\RequirePackage{ifpdf}
\ifpdf 
\documentclass[pdftex]{sigma}
\else
\documentclass{sigma}
\fi

\usepackage{stmaryrd}

\def\adots{\mathinner{\mkern1mu\raise1pt\vbox{\kern7pt\hbox{.}}\mkern2mu
 \raise4pt\hbox{.}\mkern2mu\raise7pt\hbox{.}\mkern1mu}}

\begin{document}

\allowdisplaybreaks

\renewcommand{\PaperNumber}{106}

\FirstPageHeading

\ShortArticleName{Wigner Quantization of Quadratic Hamiltonians}

\ArticleName{Wigner Quantization of Hamiltonians\\ Describing Harmonic Oscillators Coupled\\ by a General Interaction Matrix}

\Author{Gilles REGNIERS and Joris VAN DER JEUGT}

\AuthorNameForHeading{G. Regniers and J. Van der Jeugt}

\Address{Department of Applied Mathematics and Computer Science, Ghent University,\\ Krijgslaan 281-S9, B-9000 Gent, Belgium}

\Email{\href{mailto:Gilles.Regniers@UGent.be}{Gilles.Regniers@UGent.be}, \href{mailto:Joris.VanderJeugt@UGent.be}{Joris.VanderJeugt@UGent.be}}

\ArticleDates{Received September 22, 2009, in f\/inal form November 20, 2009;  Published online November 24, 2009}

\Abstract{In a system of coupled harmonic oscillators, the interaction can be represented by a real, symmetric and positive def\/inite interaction matrix. The quantization of a Hamiltonian describing such a system has been done in the canonical case. In this paper, we take a more general approach and look at the system as a Wigner quantum system. Hereby, one does not assume the canonical commutation relations, but instead one just requires the compatibility between the Hamilton and Heisenberg equations. Solutions of this problem are related to the Lie superalgebras $\mathfrak{gl}(1|n)$ and $\mathfrak{osp}(1|2n)$. We determine the spectrum of the considered Hamiltonian in specif\/ic representations of these Lie superalgebras and discuss the results in detail. We also make the connection with the well-known canonical case.}


\Classification{17B60; 17B80; 81R05; 81R12}

\section{Introduction}
In quantum as well as in classical mechanics, the harmonic oscillator is one of the most popular examples to describe harmonic movement of a particle. Their numerous applications and their analytical solvability as quantum systems explain why harmonic oscillator models are thoroughly investigated. Systems of interacting harmonic oscillators are among these well-known models. A system of $n$ one-dimensional harmonic oscillators interacting with each other can be described in its most general form as~\cite{Cramer}
\begin{gather*}
  \hat{H} = \hat{r}^\dagger   V    \hat{r}.
\end{gather*}
In this equation, $\hat{r}^\dagger$ is the vector $(\hat{p}_1^\dagger, \ldots, \hat{p}_n^\dagger,   \hat{q}_1^\dagger, \ldots, \hat{q}_n^\dagger)$, with $\hat{q}_r$ and $\hat{p}_r$ respectively the position and momentum operator of the oscillator at location~$r$, and $V$ is a positive def\/inite matrix describing the coupling in position and momentum coordinates. In the present paper, we will assume that there is no coupling involving the momentum operators. Following this approach, we can write the Hamiltonian in the following manner:
\begin{gather} \label{hamiltonian_general}
  \hat{H} =         \frac{1}{2m} \big( \hat{p}_1^\dagger \ \cdots  \ \hat{p}_n^\dagger \big)
                                 \begin{pmatrix} \hat{p}_1 \\ \vdots \\ \hat{p}_n \end{pmatrix}
              +   \frac{m}{2} \big( \hat{q}_1^\dagger \ \cdots \ \hat{q}_n^\dagger \big)
                                A
                                \begin{pmatrix} \hat{q}_1 \\ \vdots \\ \hat{q}_n \end{pmatrix}.
\end{gather}
The matrix $A$ is called the interaction matrix, and it is assumed to be real, symmetric and positive def\/inite. In order to connect the physical context of harmonic oscillators coupled by springs obeying Hooke's law to this Hamiltonian, we can rewrite $A$ as $\omega^2 I + c M$. All oscillators then have mass $m$ and natural frequency $\omega$, and the coupling constant is called $c$ ($c>0$). The $n \times n$ identity matrix is denoted $I$ and $M$ is a general real and symmetric matrix. This notation is solely introduced to be able to interpret the system physically. The essential mathematics can and will be done using the more general notation $A$.

Much interest lies in the spectrum of the Hamiltonian \eqref{hamiltonian_general}, since this yields all the possible values that might arise when measuring the energy of the system. In the standard approach for determining this spectrum, one imposes the canonical commutation relations (CCRs):
\begin{gather} \label{CCRs}
  [\hat{q}_r, \hat{q}_s] = 0, \qquad [\hat{p}_r, \hat{p}_s] = 0, \qquad [\hat{q}_r, \hat{p}_s] = i \hbar \delta_{rs}.
\end{gather}
This has been done for several types of interaction matrices in~\cite{RVdJ-09}. However, there is a more general approach to tackle this problem. Eugene Wigner was the f\/irst to realize that one does not need to assume the CCRs in order to f\/ind operators that satisfy Hamilton's equations (in operator form) and the equations of Heisenberg simultaneously. Instead, imposing that these equations are equivalent as operator equations results in a set of compatibility conditions (CCs). In a standard quantum system, these CCs are naturally satisf\/ied as a consequence of the CCRs. Wigner on the other hand questioned the fact that the relations \eqref{CCRs} can be derived from the compatibility conditions~\cite{Wigner}. His discovery that this was not the case for a single harmonic oscillator, resulted in the f\/irst Wigner quantum system~\cite{Palev-86}.

In the present paper, we consider a quantum system of $n$ coupled one-dimensional harmonic oscillators, and we treat this as a Wigner quantum system. We will demonstrate how the analysis of the spectrum of this system is connected to the Lie superalgebras $\mathfrak{osp}(1|2n)$ and $\mathfrak{gl}(1|n)$. The Wigner quantization of this string of harmonic oscillators deviates from its standard quantum mechanical counterpart by a certain parameter, corresponding to the parameter characterizing unitary irreducible representations of these Lie superalgebras. The representations that we will consider in this paper are both characterized by a parameter $p$. For $\mathfrak{osp}(1|2n)$ one f\/inds back the canonical case by choosing $p=1$.

It is interesting to note that Wigner in his original paper~\cite{Wigner} already found a connection with $\mathfrak{osp}(1|2)$, although he was unaware of this fact. His compatibility conditions were exactly the def\/ining triple relations of this Lie superalgebra, or equivalently, the triple relations of the paraboson algebra. Green~\cite{Green-1953} later generalized these results and was the f\/irst to write down the paraboson relations, connected to $\mathfrak{osp}(1|2n)$, explicitly.

Other scientists have been inspired by Wigner's approach, which has resulted in the research of several dif\/ferent Wigner quantum systems~\cite{Palev2-86, Palev-97, Blasiak-03}. Amongst them, a system of coupled harmonic oscillators with periodic boundary conditions has been studied in~\cite{LSVdJ-06}, where solutions for the position and momentum operators are found in terms of generators of the Lie superalgebra $\mathfrak{gl}(1|n)$. Analysis of the properties of this quantum system has been done in a Fock type representation space of $\mathfrak{gl}(1|n)$ and the authors found a discrete and f\/inite spectrum of the coordinate and energy operators. Another quantum system consisting of coupled harmonic oscillaters with a f\/ixed wall boundary condition has been the subject of investigation in~\cite{LSVdJ-08-1}. Here the authors present solutions in another class of representation spaces of $\mathfrak{gl}(1|n)$, called the ladder representations. These systems of coupled harmonic oscillators, however, correspond to specif\/ic interaction matrices $A$. We wish to extend the performed analysis to a general interaction matrix and investigate properties of the system in specif\/ic representations of $\mathfrak{gl}(1|n)$ and $\mathfrak{osp}(1|2n)$.

This paper contains two specif\/ic examples of interaction matrices that will get much attention throughout the manuscript. The f\/irst system is a system with constant interaction, and its Hamiltonian is
\begin{gather*}
  \hat{H}_{\rm cst}   =     \sum_{r=1}^n \left(   \frac{\hat{p}_r^2}{2m}
                                        + \frac{m \omega^2}{2} \, \hat{q}_r^2 \right)
                      + \sum_{r=0}^n \frac{cm}{2} (\hat{q}_r - \hat{q}_{r+1})^2,
\end{gather*}
with $\hat{q}_0 = \hat{q}_{n+1} \equiv 0$ (f\/ixed-wall boundary conditions). Much is known already about this system. We refer to~\cite{LSVdJ-08-1} for more information. The second example is the Hamiltonian with so-called Krawtchouk interaction. This Hamiltonian depends on a parameter $\tilde{p}$, and for $\tilde{p}=\frac{1}{2}$ one ends up with its simplest form:
\begin{gather*}
  \hat{H}_{\rm K}   =           \frac{1}{2m} \sum_{r=1}^n \hat{p}_r^2
                  +   \frac{m}{2} \sum_{r=1}^n \left( \omega^2 + \frac{c   (n-1)}{2} \right) \hat{q}_r^2
                  -  \frac{cm}{4}
                \sum_{r=1}^{n-1} \sqrt{r(n-r)}  (\hat{q}_r \hat{q}_{r+1} + \hat{q}_{r+1} \hat{q}_r).
\end{gather*}
Notice that we cannot say that $\hat{q}_r   \hat{q}_{r+1} + \hat{q}_{r+1}   \hat{q}_r = 2   \hat{q}_r   \hat{q}_{r+1}$ because we make no assumptions about the commutation relations between the position and momentum operators. In Section~\ref{sec_int_matrix} and onwards we give the interaction matrices that belong to the considered systems and explain why these examples are interesting.

In Section \ref{sec_procedure} we translate the problem into a dif\/ferent form, which is connected to Lie superalgebras. This analysis works for an interaction matrix in its most general form. We supply the reader with concrete examples of interaction matrices in Section~\ref{sec_int_matrix}. The given examples handle linear chains of coupled oscillators and analytically solvable interaction matrices. This means that the procedure from the previous section works analytically, not only in a numerical way. Section~\ref{sec_superalgebra_sol} restates the problem in terms of Lie superalgebra generators. Hereafter, one can try to determine the actual spectrum in specif\/ic representations of the Lie superalgebras $\mathfrak{gl}(1|n)$ and $\mathfrak{osp}(1|2n)$. This is done in Section~\ref{sec_spec_Vp}, where the reader is supplied with general formulae to determine the spectrum and a detailed analysis including some plots. Finally, we go back to the known canonical quantization and f\/ind connections with the previous results.

\section{The Wigner quantization procedure} \label{sec_procedure}

The Hamiltonian \eqref{hamiltonian_general} can also be written as
\begin{gather} \label{hamiltonian_ars}
  \hat{H} = \frac{1}{2m} \sum_{r=1}^n \hat{p}_r^2 + \frac{m}{2} \sum_{r,s=1}^n a_{rs} \hat{q}_r \hat{q}_s,
\end{gather}
where we have introduced the notation $a_{rs}$ for the element on position $(r,s)$ of the interaction matrix $A$. We also assume that that the position and momentum operators are self-adjoint, that is $\hat{q}_r^\dagger = \hat{q}_r$ and $\hat{p}_r^\dagger = \hat{p}_r$. Instead of imposing the canonical commutation relations \eqref{CCRs}, we just require the equivalence of Hamilton's equations (in operator form)
\begin{gather*}
  \dot{\hat{p}}_r = - \frac{\partial \hat{H}}{\partial \hat{q}_r}, \qquad
  \dot{\hat{q}}_r = \frac{\partial \hat{H}}{\partial \hat{p}_r},
\end{gather*}
and the Heisenberg equations
\begin{gather*}
  \dot{\hat{p}}_r = \frac{i}{\hbar} [\hat{H}, \hat{p}_r], \qquad
  \dot{\hat{q}}_r = \frac{i}{\hbar} [\hat{H}, \hat{q}_r].
\end{gather*}
The resulting compatibility conditions, applied for the Hamiltonian \eqref{hamiltonian_ars} become
\begin{gather} \label{CC_pq}
 [\hat{H}, \hat{q}_r]   =   - \frac{i \hbar}{m}    \hat{p}_r,        \qquad
   [\hat{H}, \hat{p}_r]  =   i \hbar m \sum_{s=1}^n a_{rs} \hat{q}_s ,
\end{gather}
with $r=1,2,\ldots,n$. We are now looking for operator solutions for $\hat{q}_r$ and $\hat{p}_r$ satisfying the compatibility conditions~\eqref{CC_pq}, with Hamiltonian \eqref{hamiltonian_ars}. Since the interaction matrix is real and symmetric, we can apply the spectral theorem to~$A$ and write
\begin{gather*}
  A = UDU^T.
\end{gather*}
In this identity, $D$ is a diagonal matrix with the real and positive eigenvalues $\mu_j$ ($j=1, \ldots, n$) of~$A$ as diagonal elements. The columns of the matrix~$U$ are the real and orthonormal eigenvectors of~$A$, $U^T$ is the transpose of $U$. Of course, $U$ is an orthonormal matrix. In other words it satisf\/ies
\begin{gather*}
  UU^T = I = U^TU.
\end{gather*}
The matrix $U$ can be used to introduce the normal coordinates and momenta. These new operators are def\/ined by
\begin{gather} \label{PQ_def}
  \begin{pmatrix} \hat{Q}_1 \\
                   \vdots   \\
                  \hat{Q}_n
  \end{pmatrix}
  = U^T   \begin{pmatrix} \hat{q}_1 \\
                           \vdots   \\
                          \hat{q}_n
          \end{pmatrix},
  \qquad
  \begin{pmatrix} \hat{P}_1 \\
                   \vdots   \\
                  \hat{P}_n
  \end{pmatrix}
  = U^T   \begin{pmatrix} \hat{p}_1 \\
                           \vdots   \\
                          \hat{p}_n
          \end{pmatrix}.
\end{gather}
The operators $\hat{P}_j$ and $\hat{Q}_j$ are self-adjoint and they do not satisfy the CCRs, just like the opera\-tors~$\hat{p}_r$ and $\hat{q}_r$. In function of the normal coordinates and momenta, the Hamiltonian \eqref{hamiltonian_general} can be rewritten as
\begin{gather*} 
  \hat{H} =         \frac{1}{2m} \big( \hat{P}_1^\dagger \ \cdots \  \hat{P}_n^\dagger \big)
                               \begin{pmatrix} \hat{P}_1 \\ \vdots \\ \hat{P}_n \end{pmatrix}
           +   \frac{m}{2} \big( \hat{Q}_1^\dagger \ \cdots \  \hat{Q}_n^\dagger \big)
                              D
                              \begin{pmatrix} \hat{Q}_1 \\ \vdots \\ \hat{Q}_n \end{pmatrix},
\end{gather*}
or, more explicitly in summation form
\begin{gather} \label{hamiltonian_PQ_sum}
  \hat{H} = \frac{1}{2m} \sum_{j=1}^n \hat{P}_j^2 + \frac{m}{2} \sum_{j=1}^n \mu_j   \hat{Q}_j^2.
\end{gather}
This Hamiltonian shows that only the eigenvalues of the interaction matrix $A$ are of essence when the system is rewritten in function of the new opera\-tors~$\hat{P}_j$ and $\hat{Q}_j$. Looking for operator solutions of the Hamiltonian~\eqref{hamiltonian_PQ_sum}, one should not forget to take the compatibility conditions \eqref{CC_pq} into account. For the normal coordinates and momenta, these translate into
\begin{gather} \label{CC_PQ}
[\hat{H}, \hat{Q}_j]  =  - \frac{i \hbar}{m}  \hat{P}_j , \qquad
 [\hat{H}, \hat{P}_j]   =  i \hbar m  \mu_j  \hat{Q}_j.
\end{gather}
This can be obtained by substituting the transformations \eqref{PQ_def} in the compatibility conditions~\eqref{CC_pq}.

It turns out that we will be able to f\/ind solutions for $\hat{Q}_j$ and $\hat{P}_j$ satisfying the CC's \eqref{CC_PQ} and the Hamiltonian in equation \eqref{hamiltonian_PQ_sum} in terms of Lie superalgebra generators. The easiest way to establish such a result, is to introduce linear combinations of the unknown operators $\hat{Q}_j$ and $\hat{P}_j$ as follows:
\begin{gather} \label{aj_def}
  a_j^{\pm} =     \sqrt{\frac{m \sqrt{\mu_j} }{2 \hbar}}   \hat{Q}_j
              \mp \frac{i}{\sqrt{2 \hbar m \sqrt{\mu_j}}}   \hat{P}_j.
\end{gather}
In terms of the operators $a_j^{\pm}$, which satisfy the adjointness relations $(a_j^\pm)^\dagger = a_j^\mp$, the Hamiltonian~\eqref{hamiltonian_PQ_sum} can be rewritten as
\begin{gather} \label{hamiltonian_aj}
  \hat{H} = \sum_{j=1}^n \frac{\hbar \sqrt{\mu_j}}{2}   \{a_j^+, a_j^-\}
          = \sum_{j=1}^n \frac{\hbar \sqrt{\mu_j}}{2}   (a_j^+ a_j^- + a_j^- a_j^+).
\end{gather}
Again, we need to have the compatibility conditions in terms of the newly introduced operators. These follow from \eqref{CC_PQ} and are
\begin{gather} \label{CC_aj}
  \bigl[ \hat{H}, a_j^\pm \bigr]
         = \pm \, \hbar \, \sqrt{\mu_j} \, a_j^\pm, \qquad j=1,2,\ldots,n.
\end{gather}
Thus we have:

\begin{theorem} \label{th_Wigner_quantization}
  The Wigner quantization of the system \eqref{hamiltonian_ars} has been reduced to the problem of finding $2n$ operators $a_j^\pm$ $(j=1, \ldots, n)$ acting in a certain Hilbert space. These operators must satisfy $(a_j^\pm)^\dagger = a_j^\mp$ and
\begin{gather} \label{defining_rel}
   \sum_{j=1}^n \big[ \sqrt{\mu_j} \, \{a_j^+, a_j^-\},   a_k^\pm \big]
                = \pm 2   \sqrt{\mu_k} \,  a_k^\pm, \qquad k=1,2,\ldots,n.
\end{gather}
The Wigner quantization procedure is reversible, so that the knowledge of the operators $a_j^\pm$ allows us to reconstruct the observables $\hat{p}_r$ and $\hat{q}_r$. The Hamiltonian is given by equation \eqref{hamiltonian_aj}.
\end{theorem}

Equation \eqref{defining_rel} is equivalent to a quantum system describing an $n$-dimensional non-isotropic oscillator~\cite[Section 2]{LVdJ-08}. For such systems, it is known that solutions in terms of Lie superalgebra generators exist~\cite{LVdJ-08}. Some specif\/ic solutions are related to the Lie superalgebras~$\mathfrak{osp}(1|2n)$ and~$\mathfrak{gl}(1|n)$, but not all solutions are known for $n>1$. We will focus on these two solutions and investigate the spectrum of our system in representations of these Lie superalgebras. However, before moving on to this analysis, we will give some explicit examples of interaction matrices.

\section{Some interaction matrices} \label{sec_int_matrix}
The Wigner quantization procedure only requires the spectral decomposition of the interaction matrix. Since we assume that $A$ is real and symmetric, the spectral theorem is always applicable. Hence, for a given interaction matrix, the Wigner quantization procedure always works as above. However, the explicit spectral decomposition has to be calculated numerically. For some matrices, we have analytically closed expressions for the eigenvalues and eigenvectors of $A$. Hamiltonians described by such interaction matrices are called \textit{analytically solvable}, and they were thoroughly investigated in~\cite{RVdJ-09}. We will consider two examples of such analytically solvable interaction matrices. The interaction of the resulting systems will be referred to as constant interaction and Krawtchouk interaction.

\textbf{Constant interaction~\cite{Cohen-77, Brun-Hartle-99, Audenaert-02}.} The interaction matrix $A_{\rm cst} = \omega^2 I + c M_{\rm cst}$, with $M_{\rm cst}$ the $n \times n$ matrix given by
\begin{gather*}
  M_{\rm cst} = \left( \begin{array}{ccccc}
                      2 & -1 &    0   &          &    \\
                     -1 &  2 &   -1   &     0    &    \\
                      0 & -1 &    2   &  \ddots  &    \\
                        &  0 & \ddots &  \ddots  & -1 \\
                        &    &        &    -1    &  2
                   \end{array} \right),
\end{gather*}
is an analytically solvable interaction matrix. It represents a linear chain of coupled identical oscillators with constant interaction throughout the chain. The corresponding Hamiltonian is
\[
  \hat{H}_{\rm cst} =   \sum_{r=1}^n \left(   \frac{\hat{p}_r^2}{2m}
                                        + \frac{m \omega^2}{2}   \hat{q}_r^2 \right)
                  + \sum_{r=0}^n \frac{cm}{2} (\hat{q}_r - \hat{q}_{r+1})^2,
\]
with $\hat{q}_0 = \hat{q}_{n+1} \equiv 0$. The interaction between the oscillators is of a nearest-neighbour type, which is a direct consequence of the tridiagonal form of the interaction matrix. The eigenvalues of $M_{\rm cst}$ and the elements $u_{ij}$ of the orthonormal matrix $U$ are given by
\begin{gather} \label{spec_decomp_cst}
  \lambda_j = 2- 2 \cos{ \left( \frac{j \pi}{n+1} \right) }, \qquad
  u_{ij} = \sqrt{ \frac{2}{n+1} } \sin{ \left( \frac{ij \pi}{n+1} \right) },
\end{gather}
with $i,j = 1, \ldots, n$. In this particular example we have
\begin{gather*}
  \mu_j = \omega^2 + 2c - 2c \cos{ \left( \frac{j \pi}{n+1} \right) }
             = \omega^2 + 4c \sin^2{ \left( \frac{j \pi}{2(n+1)} \right) }
\end{gather*}
as eigenvalues of $A_{\rm cst}$. The operators $a_j^\pm$ are constructed as shown above and the entire Wigner quantization procedure can be reproduced with the given information.

\textbf{Krawtchouk interaction~\cite{RVdJ-09}.} Another example of a matrix for which the eigenvalues and eigenvectors have analytically closed expressions, is the Krawtchouk matrix, given by
\begin{gather*}
  M_{\rm K} = \left( \begin{array}{ccccc}
                 F_0 & -E_1 &    0   &          &          \\
                -E_1 &  F_1 &  -E_2  &     0    &          \\
                  0  & -E_2 &   F_2  &  \ddots  &          \\
                     &   0  & \ddots &  \ddots  & -E_{n-1} \\
                     &      &        & -E_{n-1} &  F_{n-1}
               \end{array} \right),
\end{gather*}
where
\begin{gather*}
  E_r = \sqrt{\tilde{p}(1-\tilde{p})} \sqrt{r(n-r)}, \qquad F_r = (n-1)\tilde{p} + (1-2\tilde{p})r,
\end{gather*}
for $r=0, \ldots, n-1$. The parameter $\tilde{p}$ lies between 0 and 1. The spectral decomposition of such a matrix is known. The eigenvalues can be written as $\lambda_j = j-1$, for $j=1, \ldots, n$, and, following the notation of Section~\ref{sec_procedure}, the element at position $(i,j)$ $(i,j = 1, \ldots, n)$ of the orthonormal matrix $U$ is given by $\tilde{K}_{i-1}(j-1)$, with
\begin{gather*}
  \tilde{K}_i(j) = \left[ \binom{n-1}{i} \binom{n-1}{j} \tilde{p}^{\, i+j}(1-\tilde{p})^{n-i-j-1} \right]^{1/2}
                   \sum_{k=0}^{\min(i,j)} \frac{\binom{i}{k}\binom{j}{k}}{\binom{n-1}{k}}
                   \left( -\frac{1}{\tilde{p}} \right)^k.
\end{gather*}
These elements $\tilde{K}_i(j)$ are evaluations of normalized Krawtchouk polynomials, which explains the name of the matrix.

Consider the system \eqref{hamiltonian_general} with
\begin{gather} \label{kraw_int_matrix}
  A = \omega^2 I + c M_{\rm K}.
\end{gather}
Then this Hamiltonian is analytically solvable and can be rewritten as
\begin{gather*}
  \hat{H}_{\rm K}  =   \sum_{r=1}^n \left(   \frac{\hat{p}_r^2}{2m}
                                    + \frac{m \omega^2 + cm\tilde{p}   (n-1)}{2}   \hat{q}_r^2
                                    + \frac{cm   (1-2\tilde{p})}{2}   r \hat{q}_r^2             \right) \\
 \phantom{\hat{H}_{\rm K}  =}{}  -
                \frac{cm   \sqrt{\tilde{p}   (1-\tilde{p})}}{2}
                \sum_{r=1}^{n-1} \sqrt{r(n-r)}   (\hat{q}_r \hat{q}_{r+1} + \hat{q}_{r+1} \hat{q}_r).
\end{gather*}
It is clear that this simplif\/ies a lot if the parameter $\tilde{p}$ is chosen to be $\frac{1}{2}$. The Hamiltonian then takes the form that is presented in the introductory section.

Following the general procedure of Section~\ref{sec_procedure}, we can rewrite the problem as \eqref{defining_rel} with
\begin{gather*}
  \mu_j = \omega^2 + c \lambda_j
             = \omega^2 +c(j-1), \qquad j=1,2,\ldots,n.
\end{gather*}
The Krawtchouk interaction matrix \eqref{kraw_int_matrix} is clearly real and symmetric. Moreover, the quantities $\sqrt{\mu_j}$ are well def\/ined for all $j = 1, \ldots, n$ and the matrix \eqref{kraw_int_matrix} is positive def\/inite whenever $\omega \neq 0$.

Other types of interaction matrices connected to discrete orthogonal polynomials can be found in~\cite{RVdJ-09}. We note that the interaction matrices in the previous examples can be written as
\begin{gather} \label{int_matrix_form}
  A = \omega^2 I + cM.
\end{gather}
From now on we will always work with an interaction matrix that can be written in the form~\eqref{int_matrix_form}. As a consequence, whenever we adopt results like~\eqref{hamiltonian_aj} or~\eqref{CC_aj} from Section~\ref{sec_procedure}, one must f\/irst set $\mu_j = \omega^2 + c \lambda_j$. The entire Wigner quantization procedure described in this section can be reproduced using this transformation.

It is now the task to f\/ind operator solutions for $a_j^\pm$ that satisfy the equation
\begin{gather} \label{defining_relations}
  \sum_{j=1}^n \left[ \sqrt{\omega^2 + c \lambda_j}   \{a_j^+, a_j^-\},   a_k^\pm \right]
               = \pm 2   \sqrt{\omega^2 + c \lambda_k} \,  a_k^\pm, \qquad k=1,2,\ldots,n.
\end{gather}
As stated above, we are able to express the wanted operators in terms of generators of Lie superalgebras. We can then f\/ind solutions in specif\/ic representation spaces of these Lie superalgebras.

\section{Lie superalgebra solutions} \label{sec_superalgebra_sol}

\subsection[The $\mathfrak{gl}(1|n)$ solution]{The $\boldsymbol{\mathfrak{gl}(1|n)}$ solution}

The Lie superalgebra $\mathfrak{gl}(1|n)$ has basis elements $e_{jk}$ with $j,k = 0,1,\ldots,n$. The solutions we are about to f\/ind, will be in terms of the odd elements $e_{j0}$ and $e_{0j}$ with $j = 1,\ldots,n$. The remaining basis elements are called even elements. The odd elements have degree 1, the even elements have degree 0. The commutation and anti-commutation relations in $\mathfrak{gl}(1|n)$ are determined by the Lie superalgebra bracket
\begin{gather*}
  \llbracket e_{ij}, e_{kl} \rrbracket = \delta_{jk} e_{il} - (-1)^{\deg(e_{ij}) \deg(e_{kl})} \delta_{il} e_{kj}.
\end{gather*}
In a representation of this Lie superalgebra, the bracket $\llbracket x, y \rrbracket$ stands for an anti-commutator if~$x$ and~$y$ are both odd elements of $\mathfrak{gl}(1|n)$, and for a commutator otherwise. We can use a star condition for $\mathfrak{gl}(1|n)$ that is f\/ixed by a signature $\sigma = (\sigma_1, \ldots, \sigma_n)$, a sequence of plus or minus signs, and by
\begin{gather} \label{e0j_dagger}
  (e_{0j})^\dagger = \sigma_j   e_{j0}, \qquad j = 1,\ldots,n.
\end{gather}
We will restrict ourselves to the case where all $\sigma_j$'s are equal to $+1$, since this corresponds to the real form $\mathfrak{u}(1|n)$. In this case it is known that f\/inite-dimensional unitary representations exist~\cite{Gould}.

Solutions of \eqref{defining_relations} in terms of generators of $\mathfrak{gl}(1|n)$ are known. They have been constructed for a f\/ixed interaction matrix in~\cite{LSVdJ-06}. Therefore, it is not necessary to copy the entire analysis here. The solutions are of the form
\begin{gather} \label{sol_gl_form}
         a_j^- = \sqrt{\frac{2   |\beta_j|}{\sqrt{\omega^2 + c \lambda_j}}} \, e_{j0},
  \qquad a_j^+ = \mbox{sign}(\beta_j)
                 \sqrt{\frac{2   |\beta_j|}{\sqrt{\omega^2 + c \lambda_j}}} \, e_{0j},
  \qquad  j=1, \ldots, n ,
\end{gather}
where the $\beta_j$'s are given by ($n>1$)
\begin{gather} \label{beta_j}
  \beta_j = - \sqrt{\omega^2 + c \lambda_j}
            + \frac{1}{n-1} \sum_{k=1}^n \sqrt{\omega^2 + c \lambda_k}.
\end{gather}
It is straightforward to verify that the Hamiltonian \eqref{hamiltonian_aj} can be written as
\begin{gather} \label{Hamiltonian_LSgen_beta}
  \hat{H} = \hbar \left( \beta e_{00} + \sum_{j=1}^n \beta_j e_{jj} \right),
\end{gather}
where we have used the notation $\beta = \sum_{k=1}^n \beta_k$ to indicate that this is a constant. We want all $\sigma_j$'s in equation~\eqref{e0j_dagger} to be equal to 1, which is, together with the adjointness condition $(a_j^\pm)^\dagger = a_j^\mp$, equivalent to saying that the values $\beta_j$ need to be positive.  Examining the form of~$\beta_j$ we see that it is equal to $-\sqrt{\omega^2 + c \lambda_j}$ plus some average value of the roots of the quantities $(\omega^2 + c \lambda_j)$. Thus, it seems reasonable that half of the $\beta_j$'s will be positive and half of them will be negative. However, it is possible to prove that this is not always the case. More concretely, we will need to assume that the coupling strength $c$ is small enough. We shall refer to this as ``weak coupling''.

Note that the condition that all $\beta_j$'s are positive does not stem from the algebraic solution of~\eqref{defining_relations} by means of~ \eqref{sol_gl_form}, but only from the requirement of the star condition $(e_{0j})^\dagger = e_{j0}$, since we are primarily interested in unitary representations of the real form $\mathfrak{u}(1|n)$ of $\mathfrak{gl}(1|n)$.

\textbf{Krawtchouk interaction.} Remember that we are considering a system of $n$ harmonic oscillators that are coupled by a certain interaction matrix. Finding operator solutions for the Hamiltonian of this system treated as a WQS was proved to be equivalent to f\/inding operators $a_j^\pm$ that satisfy the relations~\eqref{defining_relations}. This equation contains the eigenvalues of $A$ and the operators~$a_j^\pm$ are dependent on the eigenvectors of~$A$. In the specif\/ic case of Krawtchouk interaction, we know that the $j$th eigenvalue $\lambda_j$ is equal to $j-1$, with $j=1,2,\ldots,n$. We will f\/ind an upper bound for the coupling strength $c$ so that in this case all the $\beta_j$'s are positive.

In order to f\/ind an upper bound for the value of $c$, we need the following property.

\begin{lemma} \label{lambda_j_helplemma}
For $C > \frac{(n-4)^2}{16}$, we have the inequality
\begin{gather*}
  \sum_{j=0}^n \sqrt{C+j}   >   (n+1)   \sqrt{C+\frac{n}{2}-1}.
\end{gather*}
Note that $C$ denotes an arbitrary positive constant in this lemma, it is not the coupling strength.
\end{lemma}

\begin{proof}
The proof goes by induction on $n$. For $n=1$ and $n=2$ the property is trivial, as one sees for example from
\[
  \sqrt{C} + \sqrt{C+1} + \sqrt{C+2}   >   3   \sqrt{C}.
\]
For larger $n$, we f\/irst notice that
\[
  \sqrt{C} + \sqrt{C+n}   >   2   \sqrt{C+\frac{n}{2}-1}
\]
if $C > (\frac{n}{4}-1)^2$. This can be verif\/ied by solving the previous inequality to $C$ as if it were an equality. Taking the square of both sides twice results in the given boundary for~$C$. Consequently, one f\/inds
\begin{gather*}
  \sum_{j=0}^n \sqrt{C+j}   =   \sqrt{C} + \sqrt{C+n} + \sum_{j=0}^{n-2} \sqrt{(C+1)+j} \\
\hphantom{\sum_{j=0}^n \sqrt{C+j}}{}  >   2   \sqrt{C+\frac{n}{2}-1} + (n-1)   \sqrt{(C+1)+\frac{n-2}{2}-1}
                           =   (n+1)   \sqrt{C+\frac{n}{2}-1},
\end{gather*}
where induction is used to justify the inequality. This is possible because if $C > \frac{(n-4)^2}{16}$, then surely $C+1 > \frac{(n-6)^2}{16}$ as long as $n \geq 1$.
\end{proof}

It is then possible to construct an upper bound for the coupling strength $c$, as is shown in the following proposition.

\begin{proposition} \label{lambda_j_propc}
Assume that the eigenvalues of the interaction matrix $A$ are equal to $\mu_j = \omega^2 + c \lambda_j$, with $\lambda_j=j-1$ $(j=1,2,\ldots,n)$. An upper bound for the coupling strength $c$ is then given by
\begin{gather*}
  c < \frac{2(2n-3) \omega^2}{(n-1)(n^2-3n+4)}.
\end{gather*}
If $c$ satisfies this condition, then all the $\beta_j$'s given in equation \eqref{beta_j} are positive.
\end{proposition}

\begin{proof}
First of all, since $\beta_j - \beta_{j-1} = \sqrt{\omega^2 + c \,(j-2)} - \sqrt{\omega^2 + c \, (j-1)} < 0$, the row $\beta_j$ ($j=1,\ldots,n$) is decreasing. All of the $\beta_j$'s will thus be positive if and only if $\beta_n$ is positive. An equivalent condition is determined by
\begin{gather} \label{beta_n_pos_j}
  \beta_n > 0 \quad \Leftrightarrow \quad
                      \sum_{j=0}^{n-1} \sqrt{\frac{\omega^2}{c} + j}
                >   (n-1)   \sqrt{\frac{\omega^2}{c} + n-1}.
\end{gather}
To prove this inequality, we want to use Lemma~\ref{lambda_j_helplemma} for $n-2$. Therefore, we need to check if the condition of the lemma is satisf\/ied. For $n \geq 2$ we have that
\[
  \frac{\omega^2}{c}   >   \frac{(n-1)(n^2-3n+4)}{2(2n-3)} \geq   \frac{(n-6)^2}{16}.
\]
Since we will only consider systems with at least two coupled harmonic oscillators ($n \geq 2$), Lemma~\ref{lambda_j_helplemma} is applicable:
\[
          \sum_{j=0}^{n-1} \sqrt{\frac{\omega^2}{c} + j}
   >   (n-1)   \sqrt{\frac{\omega^2}{c} + \frac{n}{2}-2} + \sqrt{\frac{\omega^2}{c} + n-1}.
\]
By demanding that the r.h.s.\ of the previous equation is larger than or equal to $(n{-}1)   \sqrt{\frac{\omega^2}{c} {+} n{-}1}$, we ensure that $\beta_n$ is positive. A simple calculation shows that this is true for values of $c$ that are smaller than or equal to the upper bound given in this proposition.
\end{proof}

The upper bound for $c/ \omega^2$ is of the order $4/n^2$. An idea of how accurate our approximation of the boundary value is, can be found in Table~\ref{lambda_j_table_cn}. In this table, $c_n$ denotes the highest value for the coupling strength $c$ for which $\beta_n$ and hence all the $\beta_j$'s are positive. These values can be found by solving equation~\eqref{beta_n_pos_j} numerically to $c$. The boundary value as proposed in Proposition~\ref{lambda_j_propc} is denoted by $\tilde{c}_n$.

\begin{table}[ht]\centering
\caption{Critical values $c_n/\omega^2$ in the case $\lambda_j=j-1$} \label{lambda_j_table_cn}
    \vspace{1mm}
    \begin{tabular}{c|cccc|ccc} \hline \hline
      \vspace{-0.5cm} \\
           &                &                &                         &
           &                &                &                                                     \\ [-0.75em]
       $n$ & $\frac{\tilde{c}_n}{\omega^2}$ & $\frac{c_n}{\omega^2}$ & $\frac{\tilde{c}_n}{c_n}$ & \hspace{0.5cm}
       $n$ & $\frac{\tilde{c}_n}{\omega^2}$ & $\frac{c_n}{\omega^2}$ & $\frac{\tilde{c}_n}{c_n}$   \\ [0.4em] \hline
      \vspace{-0.75cm} \\
           &                &                &                         &
           &                &                &                           \\
        4  &    0.41667     &    1.27357     &         0.32717         & \hspace{0.5cm}
        9  &    0.06466     &    0.08639     &         0.74843           \\ [0.4em]
        5  &    0.25000     &    0.51723     &         0.48334         & \hspace{0.5cm}
       10  &    0.05105     &    0.06562     &         0.77802           \\ [0.4em]
        6  &    0.16364     &    0.27857     &         0.58742         & \hspace{0.5cm}
       20  &    0.01132     &    0.01259     &         0.89893           \\ [0.4em]
        7  &    0.11458     &    0.17391     &         0.65886         & \hspace{0.5cm}
       50  &    0.00168     &    0.00175     &         0.96186           \\ [0.4em]
        8  &    0.08442     &    0.11887     &         0.71013         & \hspace{0.5cm}
       100 &    0.00041     &    0.00042     &         0.98130           \\ [0.3em] \hline \hline
    \end{tabular}
\end{table}

For example, if $n=8$, all $\beta_j$'s will be positive if the coupling strength $c < 0.11887 \, \omega^2$. The boundary value from Proposition~\ref{lambda_j_propc} is a little more pessimistic ($c < 0.08442 \, \omega^2$), but not too much. One might believe from these observations that the fraction $\tilde{c}_n / c_n$ tends to 1 for large $n$.

\textbf{Constant interaction.} The problem of f\/inding a boundary value for the coupling strength~$c$ so that all the $\beta_j$'s are positive was discussed for constant interaction in~\cite{LSVdJ-08-1}. The authors propose an estimation of the upper bound for the coupling strength. Moreover, for $n=4, \ldots, 21$ they give exact values of this upper bound~\cite[Table 1, page 22]{LSVdJ-08-1}.

\subsection[The $\mathfrak{osp}(1|2n)$ solution]{The $\boldsymbol{\mathfrak{osp}(1|2n)}$ solution}

Apart from the solution in terms of generators of the Lie superalgebra $\mathfrak{gl}(1|n)$, we can also express a class of solutions of \eqref{defining_relations} by means of $\mathfrak{osp}(1|2n)$ generators. It is known (see~\cite{Ganchev}) that this Lie superalgebra is generated by a set of $2n$ paraboson operators $b_j^\pm$ $(j=1,2,\ldots,n)$ that satisfy the relations
\begin{gather} \label{paraboson_triple}
  \bigl[ \{ b_j^\xi, b_k^\eta \}, b_l^\epsilon \bigr] =   (\epsilon - \xi) \delta_{jl} b_k^\eta
                                                        + (\epsilon - \eta) \delta_{kl} b_j^\xi.
\end{gather}
In these triple relations, $j,k$ and $l$ are elements from the set $\{1,2,\ldots,n\}$ and $\eta, \xi, \epsilon \in \{+,-\}$ (to be interpreted as $+1$ and $-1$ in the algebraic expressions $(\epsilon - \xi)$ and $(\epsilon - \eta)$). The elements~$b_j^\pm$ are the odd elements of the superalgebra, while the even elements are formed by taking anti-commutators $\{b_j^\xi, b_k^\eta\}$.

Using the relations \eqref{paraboson_triple} it is then easy to check (see also~\cite{LVdJ-08}) that the operators
\begin{gather*}
  a_j^- = b_j^-, \qquad
  a_j^+ = b_j^+,
\end{gather*}
with $j=1,2,\ldots,n$ indeed satisfy equation \eqref{defining_relations}. The Hamiltonian \eqref{hamiltonian_aj} then takes the following form:
\begin{gather}
  \hat{H}   =   \sum_{j=1}^n \frac{\hbar}{2}   \sqrt{\omega^2 + c \lambda_j} \, \{a_j^+, a_j^-\}
           =  \hbar \sum_{j=1}^n \sqrt{\omega^2 + c \lambda_j} \, h_j,       \label{Hamiltonian_osp}
\end{gather}
where we have introduced the notation $h_j = \{a_j^+, a_j^-\}/2 = \{b_j^+, b_j^-\}/2$. The Cartan subalgebra of $\mathfrak{osp}(1|2n)$ is spanned by the $n$ elements $h_j$ $(j=1,2,\ldots,n)$.

\section[The spectrum of $\hat{H}$ in a class of representations]{The spectrum of $\boldsymbol{\hat{H}}$ in a class of representations} \label{sec_spec_Vp}

In order to study the spectrum of the Hamiltonian $\hat{H}$ in terms of the $\mathfrak{gl}(1|n)$ or $\mathfrak{osp}(1|2n)$ solutions, it is necessary to work with specif\/ic representations of those Lie superalgebras. An approach to this problem with respect to the Fock-type representations $W(p)$ of $\mathfrak{gl}(1|n)$ was given in~\cite{LSVdJ-06}. Here, we will work with another type of representations $V(p)$.

\subsection[The $\mathfrak{gl}(1|n)$ representations $V(p)$]{The $\boldsymbol{\mathfrak{gl}(1|n)}$ representations $\boldsymbol{V(p)}$}

Before analyzing the spectrum of $\hat{H}$, we will summarize the main features of the representa\-tions~$V(p)$ of $\mathfrak{gl}(1|n)$. First of all, they are f\/inite-dimensional, unitary representations. For any natural number $p$, the basis vectors of~$V(p)$ are given by~\cite{KingSVdJ-06}
\begin{gather*}
  v(\theta; \textbf{r}) \equiv v(\theta; r_1, r_2, \ldots, r_n),
\end{gather*}
with $\theta \in \{0,1\}$, $r_i \in \{ 0,1,2, \ldots \}$ and $\theta + r_1 + \dots + r_n = p$. The dimension of the vector space~$V(p)$ equals
\begin{gather*}
  \binom{p+n-1}{n-1} + \binom{p+n-2}{n-1},
\end{gather*}
in which both terms represent the number of basis vectors for $\theta=0$ and $\theta=1$ respectively. The action of the $\mathfrak{gl}(1|n)$ generators on these basis vectors can be determined~\cite{KingSVdJ-06}. We will give the actions of the diagonal elements $e_{00}$ and $e_{kk}$ since only these actions will be needed to f\/ind the spectrum of the Hamiltonian:
\begin{gather*}
  e_{00} v(\theta; \textbf{r})   =   \theta   v(\theta; \textbf{r}), \\
  e_{kk} v(\theta; \textbf{r})   =   r_k \, v(\theta; \textbf{r}).
\end{gather*}
In terms of the $\mathfrak{gl}(1|n)$ generators, the Hamiltonian takes the form \eqref{Hamiltonian_LSgen_beta}:
\[
  \hat{H} = \hbar \left( \beta e_{00} + \sum_{j=1}^n \beta_j e_{jj} \right).
\]
Clearly, looking at the actions of the elements $e_{00}$ and $e_{kk}$, the vectors $v(\theta; \textbf{r})$ are eigenvectors for~$\hat{H}$:
\[
  \hat{H} v(\theta; \textbf{r}) = \hbar E_\textbf{r}  v(\theta; \textbf{r}),
\]
where the eigenvalues $\hbar E_\textbf{r}$ are determined by
\begin{gather} \label{spec_gl1n_Er}
  E_\textbf{r} = \beta p - \sum_{j=1}^n \sqrt{\omega^2 + c \lambda_j} \, r_j.
\end{gather}
This can be established by noting that
\[
  \beta = \sum_{j=1}^n \beta_j = \frac{1}{n-1} \sum_{j=1}^n \sqrt{\omega^2 + c \lambda_j}
\]
and by using the fact that $\theta + r_1 + \dots + r_n = p$.

In the case where there is no coupling $(c=0)$, all the $\beta_j$'s become the same. It follows that $\beta_j = \omega/(n-1)$ and $\beta = n \omega/(n-1)$. In this case, we thus see that the eigenvalues of $\hat{H}$ are
\[
  \hbar \omega \left(\frac{p}{n-1} + \theta\right).
\]
So in fact, there are two eigenvalues. The lowest one, for $\theta=0$, has multiplicity $\binom{p+n-1}{n-1}$. The highest eigenvalue has multiplicity $\binom{p+n-2}{n-1}$.

\begin{figure}[htb]\centering
\begin{minipage}{70mm}\centering
   {\small Constant interaction}   \\[-3mm]
 \includegraphics[angle=270, width=60mm]{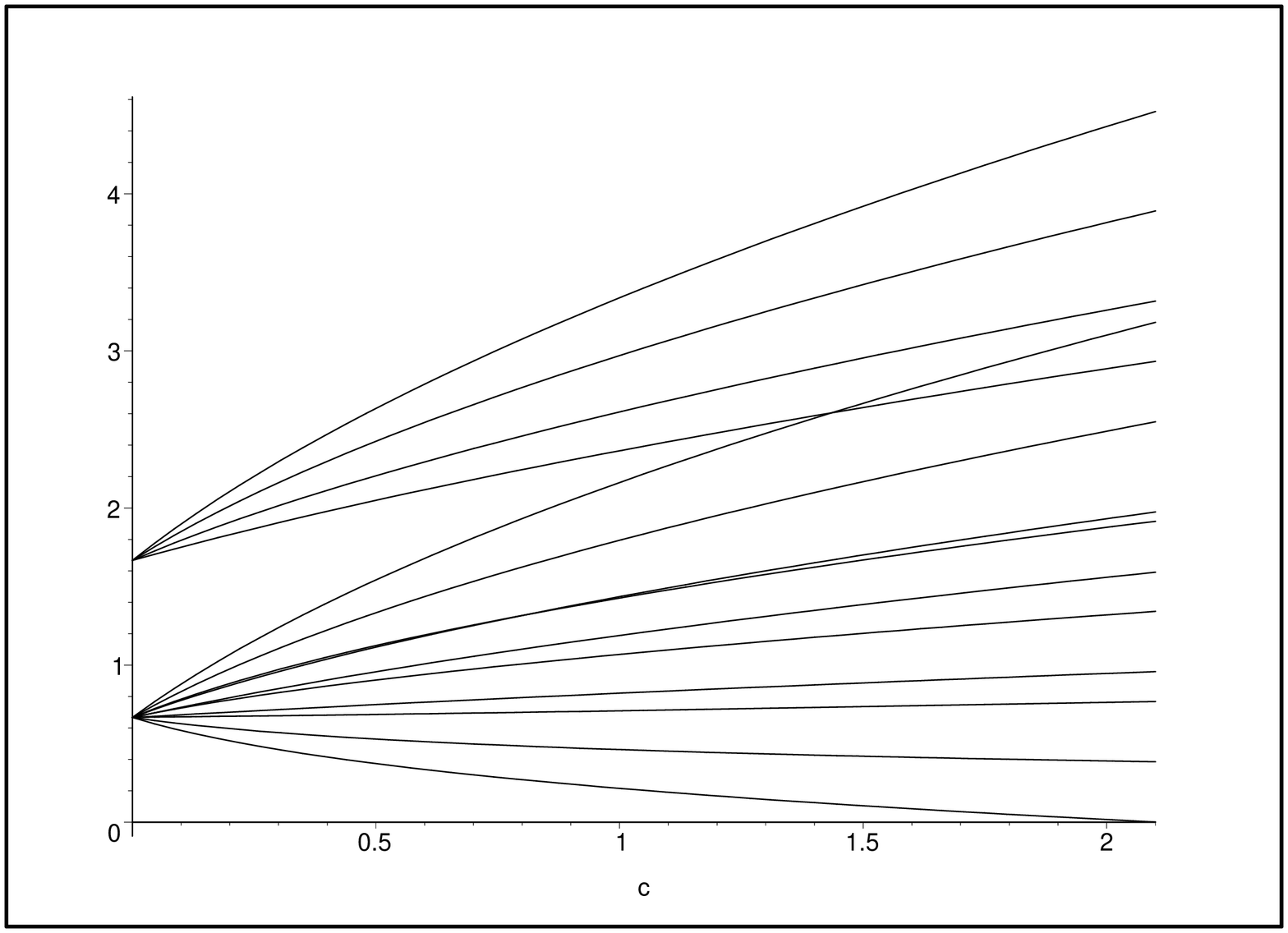}                                                       \end{minipage}
 \
 \begin{minipage}{70mm}\centering
   {\small Krawtchouk interaction}    \\[-3mm]
    \includegraphics[angle=270, width=60mm]{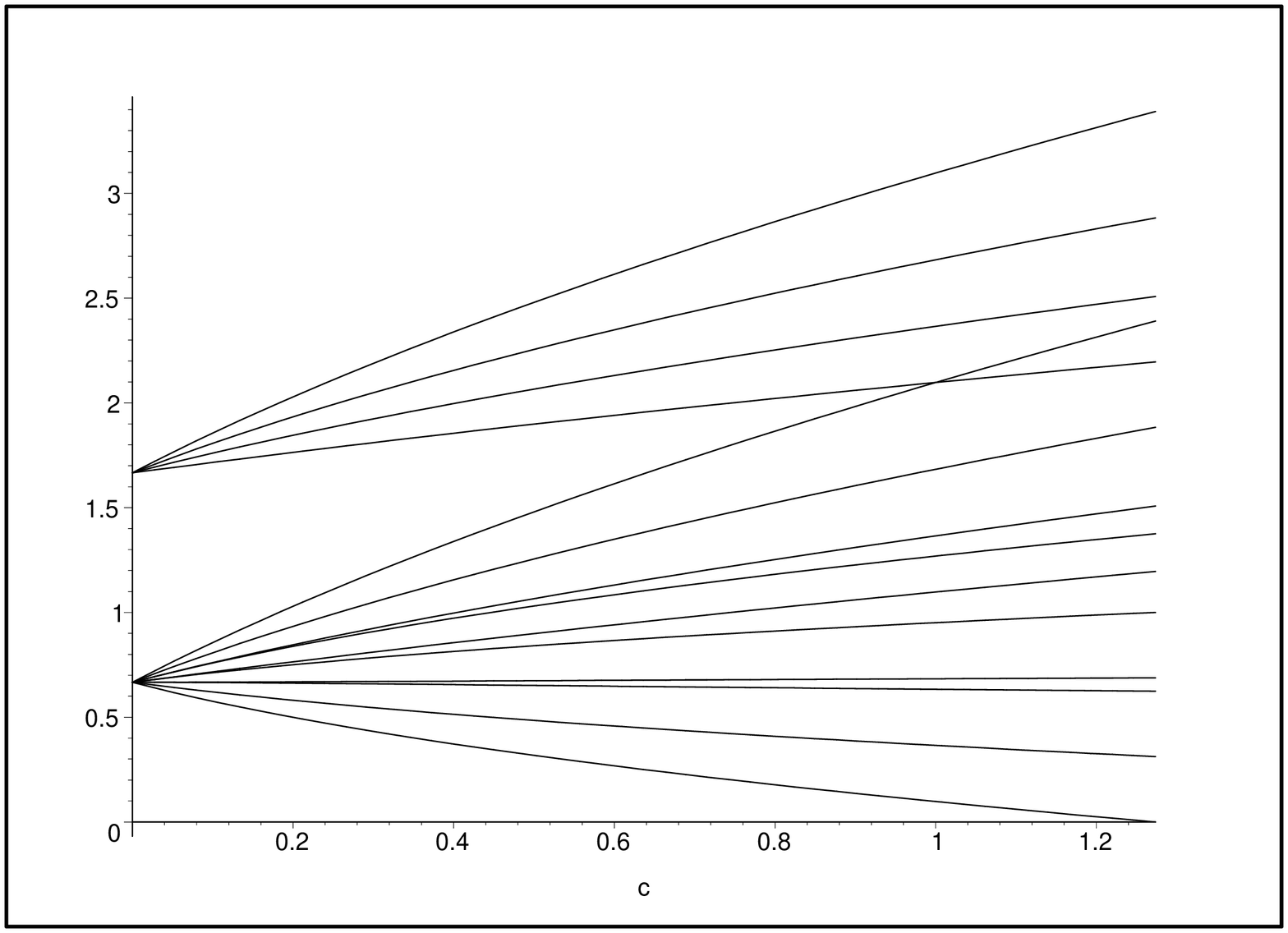}
  \end{minipage}

\caption{Spectrum of the Hamiltonian \eqref{Hamiltonian_LSgen_beta} in the $\mathfrak{gl}(1|n)$ representation $V(p)$ for $n=4$, $p=2$ and $\omega=\hbar=1$, as a function of the coupling constant~$c$. The left f\/igure belongs to the system with constant interaction, where the $\lambda_j$ are given by equation \eqref{spec_decomp_cst}. The right f\/igure represents the Krawtchouk case, with $\lambda_j = j-1$.}
\label{fig_spec_gl}
\end{figure}

Our main interest lies in the weak coupling case, where $0< c <c_n$. The energy levels are easily computed through equation \eqref{spec_gl1n_Er}. The result for $n=4, p=2$ and $\omega=\hbar=1$ can be seen in Fig.~\ref{fig_spec_gl}, where we have chosen to compare the systems with constant and Krawtchouk interaction.

Both f\/igures look quite similar, but there are some dif\/ferences. We see that in general all eigenvalues are dif\/ferent, but for specif\/ic values of~$c$ some of the energy levels cross each other. For these values of~$c$, the multiplicity of some of the eigenvalues is higher than~$1$. In the constant interaction case, we see that energy levels can cross if we restrict ourselves to, say, $\theta=0$. Also, note that there are indeed only two eigenvalues in the case without coupling ($c=0$).

The f\/igure also suggests that the lowest energy level tends to zero as the coupling strength reaches $c_n$. In order to prove this, we need to know the lowest energy level. First, we note that $\beta_n \leq \beta_j$ for all $j$ as soon as $\lambda_n \geq \lambda_j$ for all $j$. We can always choose $\lambda_n$ to be the largest eigenvalue, so we can assume that $\beta_n$ is the smallest of all $\beta_j$. Next, the formula
\[
  E_\textbf{r} = \beta \theta + \sum_{j=1}^n \beta_j r_j
\]
gives the energy levels as a sum of $p$ terms. Thus, the lowest energy level arises when all $p$ terms are equal to $\beta_n$. The def\/inition of $c_n$ tells us that $\beta_n=0$ for $c=c_n$, so the lowest energy level~$p \beta_n$ tends to zero when $c$ approaches $c_n$.

\subsection[The $\mathfrak{osp}(1|2n)$ representations $V(p)$]{The $\boldsymbol{\mathfrak{osp}(1|2n)}$ representations $\boldsymbol{V(p)}$}

We will also take a look at the (inf\/inite-dimensional) representations $V(p)$ of $\mathfrak{osp}(1|2n)$, with lowest weight $(\frac{p}{2}, \ldots, \frac{p}{2})$. Such a representation is a unitary, irreducible representation (unirrep) if and only if $p \in \{ 1,2,\ldots,n-1 \}$ or $p>n-1$ \cite[Theorem~7]{LSVdJ-08-2}. In literature, where $\mathfrak{osp}(1|2n)$ is related to the $n$-paraboson algebra, the parameter $p$ is sometimes referred to as the order of the parastatistics. A basis for the representations $V(p)$ was given in~\cite{LSVdJ-08-2}, and consists of all Gelfand--Zetlin patterns for partitions of length at most $n$. These GZ-vectors have the following form:
\begin{gather*} 
  |m) \equiv |m)^n
      \equiv \left| \begin{array}{cccc}
                        m_{1 \, n}  & \cdots &  m_{n-1, \, n}  & m_{nn} \\
                      m_{1, \, n-1} & \cdots & m_{n-1, \, n-1} &        \\
                          \vdots    & \adots &                 &        \\
                          m_{11}    &        &                 &
                    \end{array}
             \right)
      \equiv \left| \begin{array}{l}
                        [m]^n   \\
                      |m)^{n-1}
                    \end{array}
             \right).
\end{gather*}
The top line of the GZ-vectors is any partition into at most $p$ parts, where $p$ is the label of the representation. This partition is denoted by $[m]^n$. For basic information about partitions, we refer to~\cite{MacDonald}. The other elements of the GZ-vectors, denoted by $|m)^{n-1}$ satisfy the \textit{betweenness conditions}
\begin{gather*}
  m_{i,   j+1} \geq m_{ij} \geq m_{i+1,   j+1}, \qquad  1 \leq i \leq j \leq n-1.
\end{gather*}
The actions of the $\mathfrak{osp}(1|2n)$ generators on these basis vectors are known~\cite{LSVdJ-08-2}. In particular, the action of the diagonal elements $h_j$ is given by
\begin{gather*}
  h_j   |m) = \left( \frac{p}{2} + \sum_{r=1}^j m_{rj} - \sum_{r=1}^{j-1} m_{r,   j-1} \right) |m).
\end{gather*}
The Hamiltonian of the system in terms of $\mathfrak{osp}(1|2n)$ generators was given by equation \eqref{Hamiltonian_osp}:
\[
  \hat{H} = \hbar \sum_{j=1}^n \sqrt{\omega^2 + c \lambda_j} \, h_j.
\]
From the action of the diagonal elements $h_j$ it is clear that the vectors $|m)$ are eigenvectors of the Hamiltonian. We can write
\[
  \hat{H}   |m) = \hbar   E_m   |m),
\]
in which $E_m$ stands for
\begin{gather} \label{osp_eigenv_Em}
  E_m = \sum_{j=1}^n \sqrt{\omega^2 + c \lambda_j}
        \left( \frac{p}{2} + \sum_{r=1}^j m_{rj} - \sum_{r=1}^{j-1} m_{r,   j-1} \right).
\end{gather}
In the case without coupling $(c=0)$ we see that the eigenvalues simplify signif\/icantly and they can be written in the form
\[
  \hbar   \omega \left( \frac{np}{2} + \sum_{r=1}^n m_{rn} \right).
\]
The summation in this expression is in fact the weight of the partition $[m]^n$. This weight can be any positive integer~$k$, which we shall call the height of the eigenvalue $E_k^{(p)}$. This means that for $c=0$ there is an inf\/inite amount of eigenvalues, that can be written as
\begin{gather*}
  E_k^{(p)} = \hbar   \omega \left( \frac{np}{2} + k \right), \qquad k = 0,1,2, \ldots.
\end{gather*}
The multiplicity $\mu(E_k^{(p)})$ of each eigenvalue can be determined with the help of some theoretical arguments. First of all, $\mu(E_k^{(p)})$ will be equal to the total number of GZ-vectors with a partition~$\nu$ in the top row, where $\nu$ is any partition of $k$ into at most $p$ parts. Let $\nu'$ be the conjugate partition of $\nu$~\cite{MacDonald}. It is known (see for example~\cite[Section 4.6]{Wybourne-1970}) that the representation of $\mathfrak{gl}(n)$ that is labelled by the partition $\nu$ has dimension $\binom{n}{\nu'}$, where we have used the generalization of the binomial coef\/f\/icient for a partition~\cite[page~45]{MacDonald}. This is def\/ined by
\begin{gather*}
  \binom{X}{\nu} = \prod_{(i,j) \in \nu} \frac{X - c(i,j)}{h(i,j)},
\end{gather*}
where $c(i,j) = j-i$ and $h(i,j) = \nu_i + \nu_j' -i-j+1$ are the content and the hook length of $(i,j)$ respectively. So for a given partition $\nu$, the number of GZ-patterns that have $\nu$ in the top row equals $\binom{n}{\nu'}$. This implies that the multiplicity of each eigenvalue is equal to
\begin{gather} \label{osp_mult_eigenv_c0}
  \mu(E_k^{(p)}) = \sum_{\nu, \, |\nu|=k, \, l(\nu) \leq \lceil p \rceil} \binom{n}{\nu'}.
\end{gather}
The ceiling function $\lceil p \rceil$ is used to cover the cases where $n-1 < p < n$. So we have found that the energy levels for $c=0$ are equidistant with spacing $\hbar \, \omega$ and the multiplicities of the eigenvalues can be computed through equation \eqref{osp_mult_eigenv_c0}.

In the case with actual coupling $(c \neq 0)$ the eigenvalues can be found by equation \eqref{osp_eigenv_Em}. Unlike the weak coupling case in the representations $V(p)$ of $\mathfrak{gl}(1|n)$, the multiplicities of the eigenvalues are not all equal to one. Any two basis vectors $|m)$ and $|m')$ that are subject to
\begin{gather} \label{condition_same_Em}
  \sum_{r=1}^j m_{rj} = \sum_{r=1}^j m_{rj}' \qquad \forall \; j=1,2, \ldots, n
\end{gather}
yield the same eigenvalue $\hbar E_m$. So the multiplicity of an eigenvalue $\hbar E_m$ is equal to the number of basis vectors for which the sum of the elements on row $j$ is equal to the sum of the elements on row $j$ of $|m)$, for every~$j$. For example, the vectors
\[
  |m) \equiv \left| \begin{array}{cccc}
                      5 & 0 & 0 & 0 \\
                      4 & 0 & 0 &   \\
                      2 & 0 &   &   \\
                      1 &   &   &
                    \end{array}
             \right)
  \mbox{\qquad and \qquad}
  |m') \equiv \left| \begin{array}{cccc}
                       3 & 2 & 0 & 0 \\
                       3 & 1 & 0 &   \\
                       2 & 0 &   &   \\
                       1 &   &   &
                     \end{array}
              \right)
\]
yield the same eigenvalue. From this it is also clear that the total number of distinct eigenvalues at height $k$ is equal to that number for $p=1$. Indeed, every vector with~$p>1$ can be associated with a vector with $p=1$ for which equation~\eqref{condition_same_Em} holds, as can be seen in the previous example. Moreover, all eigenvalues in the case $p=1$ clearly have multiplicity~1 (in the generic case when all~$\lambda_j$ are distinct). We know what the number of eigenvalues at height $k$ for $p=1$ is, namely
\[
    \sum_{\nu, \, |\nu|=k, \, l(\nu) \leq 1} \binom{n}{\nu'}
  = \binom{n}{(k)'}
  = \prod_{j=1}^k \frac{n+j-1}{j},
\]
where the default value for $k=0$ is equal to 1. The latter product is nothing more than the binomial coef\/f\/icient $\binom{n+k-1}{n-1}$, which shows that it is an integer.

\begin{figure}[htb]\centering
\begin{minipage}{70mm}\centering
   {\small      $k=1$ }\\[-3mm]
   \includegraphics[angle=270, width=60mm]{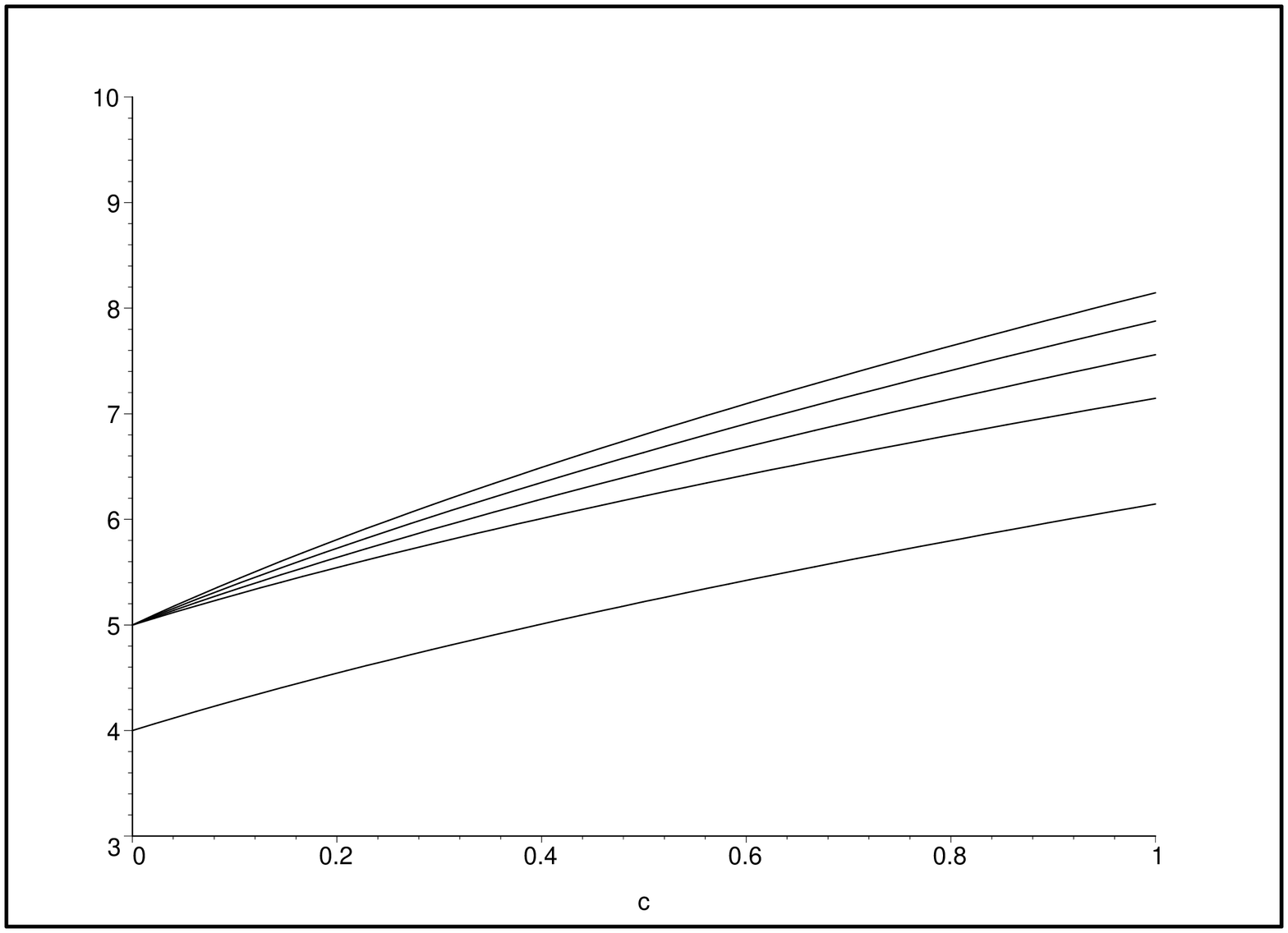}
   \end{minipage}
   \
\begin{minipage}{70mm}\centering
   {\small   $k=2$ }\\[-3mm]
    \includegraphics[angle=270, width=60mm]{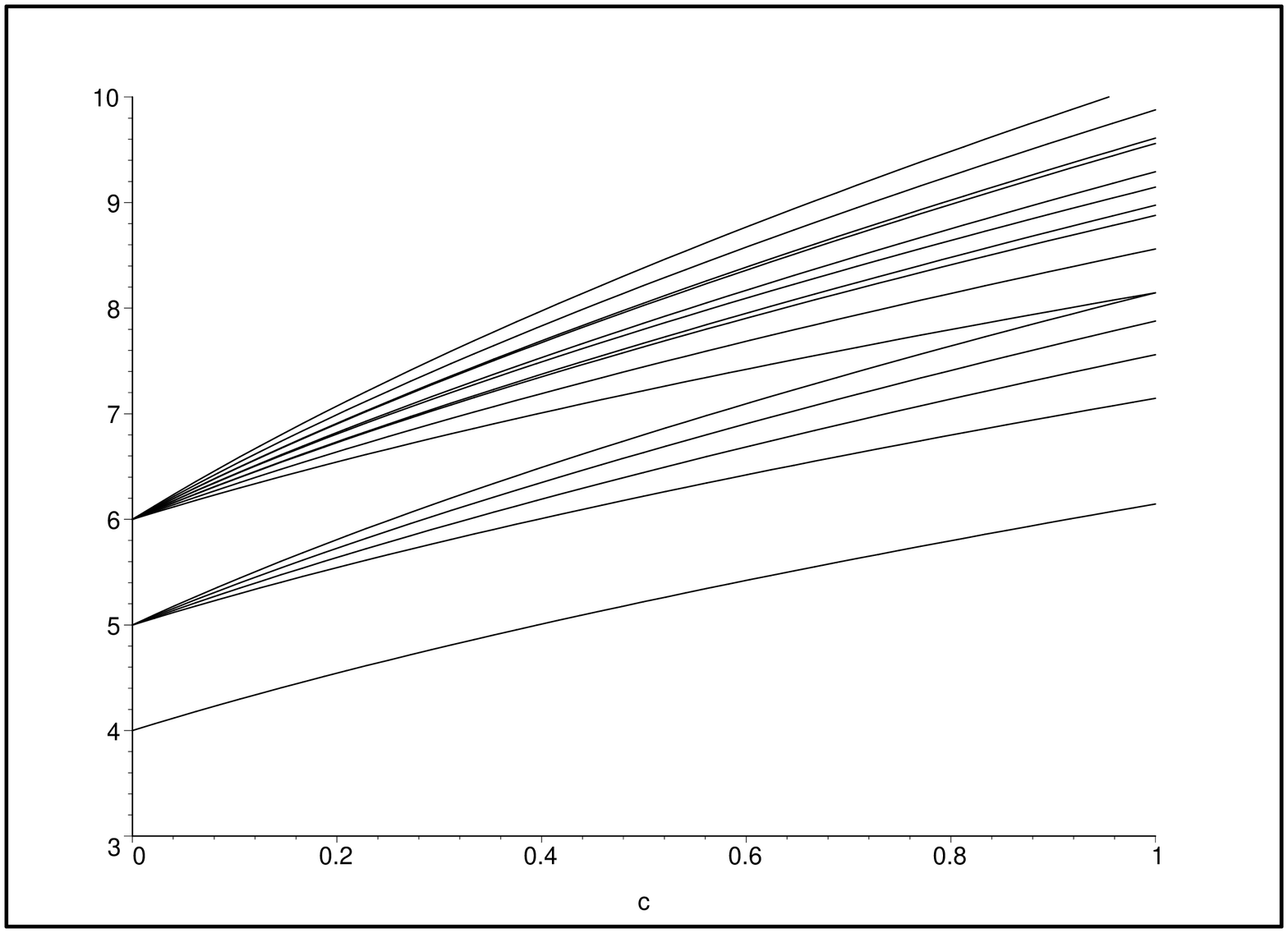}
  \end{minipage}

\caption{Spectrum of the Hamiltonian \eqref{Hamiltonian_osp} in the $\mathfrak{osp}(1|2n)$ representation $V(p)$ for $n=4$, $p=2$, $\omega=\hbar=1$ and $\lambda_j = j-1$, as a function of the coupling constant~$c$. The left f\/igure gives the spectrum up to height~$k=1$, the image on the right goes one step higher~($k=2$). The total spectrum is inf\/inite.}
\label{fig_spec_osp}
\end{figure}

It is now clear that some eigenvalues have multiplicity greater than 1. Furthermore it is possible that some of the energy levels cross each other, just as in the $\mathfrak{gl}(1|n)$ case. This means that for specif\/ic values of $c$ there are some eigenvalues for which the multiplicity is even higher. It would be inappropriate to try to compute these values of $c$. Let us instead look at Fig.~\ref{fig_spec_osp}, where we have plotted a part of the energy spectrum for $n=4, p=2$, $\omega=\hbar=1$ and $\lambda_j = j-1$, to visualize things.  Recall that we are dealing with an inf\/inite spectrum. Therefore, we will only plot the spectrum up to height $k$, for $k=1$ and $k=2$.

The eigenvalues on height 0 and 1 all have multiplicity~1 for $c>0$ and they never cross. The f\/igure on the right shows the energy values for $k=2$ as well, where we both have higher multiplicities and crossing energy levels. Six of the ten distinct energy levels at height~2 have multiplicity~2.

\section{Relation to canonical quantization} \label{sec_canonical_case}
Having treated the quantum mechanical system \eqref{hamiltonian_general} as a Wigner quantum system in general, it is interesting to consider the canonical case as a special case. In this case, one assumes the canonical commutation relations (CCRs)
\begin{gather} \label{CCR_q}
  \bigl[ \hat{q}_r, \hat{q}_s \bigr] = 0, \qquad
  \bigl[ \hat{p}_r, \hat{p}_s \bigr] = 0, \qquad
  \bigl[ \hat{q}_r, \hat{p}_s \bigr] = i \hbar \delta_{rs}.
\end{gather}
Similarly as before, we write $A = U^T D U$, where $U$ is the orthonormal matrix with the eigenvectors of $A$ as rows and $D$ is the diagonal matrix with the eigenvalues of $A$ on the main diagonal. We def\/ine new operators $\hat{Q} = U (\hat{q}_1 \ldots \hat{q}_n)^T$ and $\hat{P} = U (\hat{p}_1 \ldots \hat{p}_n)^T$. These are subject to the same commutation relations as in equations \eqref{CCR_q} and yield a new expression for the Hamiltonian:
\[
  \hat{H} = \sum_{j=1}^n \left( \frac{1}{2m} \hat{P}_j \hat{P}_j^T
                                + \frac{m}{2} (\omega^2 + c \lambda_j) \hat{Q}_j \hat{Q}_j^T \right).
\]
We then take the linear combinations \eqref{aj_def} of the operators $\hat{Q}$ and $\hat{P}$, and thus create new opera\-tors~$a_j^\pm$. The Hamiltonian can be written in terms of these new operators:
\[
  \hat{H} = \sum_{j=1}^n \frac{\hbar}{2}   \sqrt{\omega^2 + c \lambda_j} \, \{a_j^+, a_j^-\}.
\]
Using the canonical commutation relations of $\hat{Q}$ and $\hat{P}$, one f\/inds that the operators $a_j^\pm$ satisfy the usual boson commutation relations:
\[
  \bigl[ a_j^\pm, a_k^\pm \bigr] = 0, \qquad \bigl[ a_j^-, a_k^+ \bigr] = \delta_{jk}.
\]
As before, it can also be verif\/ied that
\begin{gather} \label{CC_can}
  \bigl[ \hat{H}, a_j^\pm \bigr]
         = \pm   \hbar   \sqrt{\omega^2 + c \lambda_j} \, a_j^\pm, \qquad  j=1,2,\ldots,n.
\end{gather}
These are in fact the compatibility conditions derived in the general case, interpreting the system as a WQS. These CCs are also valid in the canonical case, since the CCRs imply the CCs.

We will now def\/ine the $n$-boson Fock space, which is equivalent to the representation $V(1)$ of~$\mathfrak{osp}(1|2n)$. Since $p=1$ represents the canonical case, we will f\/ind a correspondence between the basis vectors of $V(1)$ and the basis vectors of the $n$-boson Fock space. The latter are constructed from a vacuum vector $| 0 \rangle$, with
\[
  \langle 0|0 \rangle = 1, \qquad a_j^-  | 0 \rangle = 0.
\]
The other (orthogonal and normalized) basis vectors are then def\/ined by
\begin{gather} \label{Fock_space_basis_vectors}
   | k_1, \ldots, k_n \rangle = \frac{ (a_1^+)^{k_1} \cdots (a_n^+)^{k_n} }
                                         { \sqrt{k_1! \cdots k_n!} }
                                     | 0 \rangle,
\end{gather}
where $k_1, \ldots, k_n \in \mathbb{Z}_+$. We need to f\/ind a correspondence between these vectors and the Gelfand--Zetlin basis vectors of the representation $V(1)$ of $\mathfrak{osp}(1|2n)$, generally denoted by
\begin{gather} \label{GZ_p1}
  \left| \begin{array}{ccccc}
              m_n   &    0   & \cdots & 0 & 0 \\
            m_{n-1} &    0   & \cdots & 0 &   \\
            \vdots  & \vdots & \adots &   &   \\
              m_2   &    0   &        &   &   \\
              m_1   &        &        &   &
         \end{array}
  \right).
\end{gather}
Using $\bigl[ a_j^-, a_j^+ \bigr] = 1$, one f\/inds that
\[
  \hat{H}  | 0 \rangle = \frac{\hbar}{2} \sum_{j=1}^n \sqrt{\omega^2 + c \lambda_j}  | 0 \rangle
                           = \hbar E_0  | 0 \rangle,
\]
where we have used the notation $E_0$ to indicate the lowest energy level. This is the lowest energy state of the system. The higher energy levels can be calculated using equation~\eqref{CC_can} in a~straightforward way. This results in
\begin{gather*}
        \hat{H}  | k_1, \ldots, k_n  \rangle
    =   \hbar \left( E_0 + \sum_{j=1}^n k_j \sqrt{\omega^2 + c \lambda_j} \right)
              | k_1, \ldots, k_n  \rangle.
\end{gather*}
By comparison with equation \eqref{osp_eigenv_Em} for $p=1$, one f\/inds that
\begin{gather*}
  k_j   =   \sum_{r=1}^j m_{rj} - \sum_{r=1}^{j-1} m_{r,   j-1}  =  m_j - m_{j-1}.
\end{gather*}

Thus we have:

\begin{proposition}
  The $n$-boson Fock space and the $\mathfrak{osp}(1|2n)$ representation space $V(1)$ are equiva\-lent and their basis vectors~\eqref{Fock_space_basis_vectors} and~\eqref{GZ_p1} are related by $k_j = m_j - m_{j-1}$.
\end{proposition}

\section{Conclusion}

To conclude, we have in this paper considered the quantization of a system of harmonic oscillators as a Wigner quantum system. The quadratic coupling terms have been characterized by an interaction matrix. For such systems, the Wigner quantization procedure can be performed completely (Theorem~\ref{th_Wigner_quantization}), leading to a set of algebraic triple relations~\eqref{defining_rel} as compatibility conditions. These relations have particular solutions in terms of generators of the Lie superalgeb\-ras~$\mathfrak{gl}(1|n)$ or~$\mathfrak{osp}(1|2n)$. Then the unitary representations of these Lie superalgebras play an important role: the algebraic generators, and thus also the physical operators corresponding to observables, act in these representations. For some classes of representations, the spectrum of the Hamiltonian operator is determined explicitly, and discussed.

As leading examples throughout the paper, we consider two analytically solvable systems. The f\/irst is a classical one, describing a linear chain of harmonic oscillators coupled by a harmonic nearest-neighbour interaction. The second is a relatively new system, again describing a chain of harmonic oscillators, but this time the nearest-neighbour coupling is a ``Krawtchouk coup\-ling''.

The original results in this paper are: the proof that all Hamiltonians with quadratic interaction terms can be reduced to an algebraic set of relations under Wigner quantization (Section~\ref{sec_procedure}); the conditions for the coupling strength in the case of Krawtchouk interaction for the $\mathfrak{gl}(1|n)$ solution (Section~\ref{sec_superalgebra_sol}); the determination and discussion of the energy spectrum for specif\/ic representations of the Lie superalgebras $\mathfrak{gl}(1|n)$ and $\mathfrak{osp}(1|2n)$ (Section~\ref{sec_spec_Vp}). Wigner quantization is an extension of canonical quantization, so the canonical case appears as one particular solution of the various solutions allowed by Wigner quantization. For the systems under consideration here, the canonical case corresponds to the representation $V(1)$ of the $\mathfrak{osp}(1|2n)$ solution. This correspondence is explained in Section~\ref{sec_canonical_case}.

In this paper we have dealt with the Wigner quantization of Hamiltonians of the form~\eqref{hamiltonian_ars}, where the interaction terms are explicitly present, so there is `dynamical interaction'. It is interesting to note that Wigner quantization also allows another peculiar feature. In~\cite{Palev-06}, one considers the Wigner quantization of a system of free oscillators (i.e.\ the Hamiltonian contains no dynamical interaction terms), and it is shown that certain solutions of the compatibility conditions can be interpreted as being responsible for statistical interactions in the system. So in such a case the purely algebraic quantization condition implies a statistical interaction in a~free system.

\subsection*{Acknowledgments}

G.\ Regniers was supported by project P6/02 of the Interuniversity Attraction Poles Programme (Belgian State --
Belgian Science Policy).

\pdfbookmark[1]{References}{ref}

\LastPageEnding

\end{document}